\documentclass[12pt,a4paper]{article}

\usepackage{epsfig,amstex}
\usepackage{amssymb}
\usepackage{graphics}
\begin{document}
\vspace*{-1.8cm}
\begin{flushright}{\bf LAL 01-71}\\
{October 2001}\\
\end{flushright}
\vskip 1.5 cm

\begin{center}
{\bf\LARGE
Possible use of the dedicated MARLY\\
one meter telescope for intensive\\
\vspace*{0.2cm}
supernov{\ae} studies} 
\end{center}

\vspace*{0.5cm} 
\begin{center}
{\large\bf  M. Moniez} {(moniez@@lal.in2p3.fr)},\\
{\large\bf  O. Perdereau} (perderos@@lal.in2p3.fr)
\end{center}

\begin{center}
\large {\bf 
Laboratoire de l'Acc\'el\'erateur Lin\'eaire}\\
\normalsize {
CNRS-IN2P3 et Universit\'e Paris-Sud, B\^at. 200, BP 34 - 91898 Orsay cedex}
\end{center}
\section{Introduction}

EROS has discovered $\sim 70$ SNs during 8 periods partially dedicated
to a SN search. The average discovery rate is 1 SN per 2 hours of
observation, corresponding to expectation. Before 1999, the SNs that we
discovered were publicly available on the web and reported in IAU
telegrams. The low scientific return of this procedure convinced the EROS
collaboration to work within a larger collaboration.
Nevertheless, EROS has taken advantage of the homogeneity of its
SNIa sample to obtain a SNIa explosion rate at
$z\sim 0.1$\cite{tauxSN}.

EROS has participated in the large campaign of February-March 1999
lead by the SCP, which aimed at measuring $\sim 10$ SNIa's discovered
before or at maximum, including photometric and spectroscopic follow-up.
EROS contributed an homogeneous sample of 8 SNIa's discovered
before maximum (of which 7 were monitored).
Our procedure was as follows:
The first SN detection was made on the day following the night during
which 20 pictures (one square degree each) were taken.
A confirmation picture was taken the following night, and
the SN coordinates were then given to the coordinator.
The complete process used to take
less than 48 hours to provide a {\it confirmed} SN.
This delay was clearly acceptable, if one considers the $\gtrsim 50\% $ proportion
of pre-maximum SNIa's delivered by EROS.
A preliminary analysis of the photometric data collected for these SNIa's
is described in detail in Nicolas Regnault's thesis\cite{regnault}.

The EROS-2 microlensing search will end at the end of 2002. 
In this document,
we investigate a new  way of using the EROS telescope
(the MARLY) after this date. 

\section{The MARLY as a SN-photometer}
\subsection{Today's status}
The MARLY present optics has a wide aperture, and the sampling is relatively
poor (EROS was designed to maximise the number of monitored stars per
pixel). The pixel size is 0.6 arcsec.
A test, done with the heat sources of the dome switched off,
showed that the dome contribution to the experimental 
seeing was 0.8 arcsec on the optical axis.
As there is a significant degradation of the seeing 
off-axis 
and because of the effect of the heat sources,
the median instrumental seeing is only 1.5 arcsec.

\subsection{Telescope configuration changes}
The Marly telescope had originally a F/8 aperture. For the EROS programme,
a focal reducer has increased the aperture to F/5. With the original
aperture, a simple setup with a one-CCD camera and a {\bf BVRI} 
filter wheel could
be installed on the telescope.\\
We will also propose a more sophisticated setup to perform {\bf U} band 
photometry in parallel with the other bands. 

\subsection{Optical photometer performances}
In this section, we estimate the total time needed
to fully monitor a SNIa with the MARLY telescope
every other night in {\bf BVRI}
during 80 days after its discovery.
\subsubsection{Expression of the exposure time}
For a given passband $w$,
the exposure time $T_{exp}$ needed to reach a relative photometric
precision $\delta$ on a pointlike object with apparent magnitude $M_w$
depends on
\begin{itemize}
\item
the photoelectron flux  $\Phi_w(M_w=0)$ (in $\gamma e^-/s/m^2$) associated
with a source of magnitude $M_w=0$,
taking into account the detector efficiency,
\item
the magnitude of the background surface brightness $\mu_w$
per $arcsec^2$,
\item
the seeing.
\end{itemize}
For the MARLY collecting surface,
the calculation and the validity conditions of the following
formula are given in annex A:
\begin{equation}
\begin{split}
T_{exp}
& \simeq \frac{(175 s)}{\left[\delta(in \%)\right]^2}
\left[\frac{\Phi_w(M_w=0)}{10^9}\right]^{-1}\times	\\
& \quad \times 10^{\frac{M_w-17.5}{2.5}}
\left[0.27+1.59\sqrt{0.22+\left[\frac{seeing}{1\ arcsec}\right]^2
10^{\frac{M_w-\mu_w}{2.5}}}\right]^2
\end{split}
\end{equation}

We will use this expression to perform a realistic simulation,
taking into account real distributions for each parameter.
\subsubsection{Photoelectron signal expected from the SNIa's}
\begin{figure}
\begin{center}
\mbox{\epsfig{file=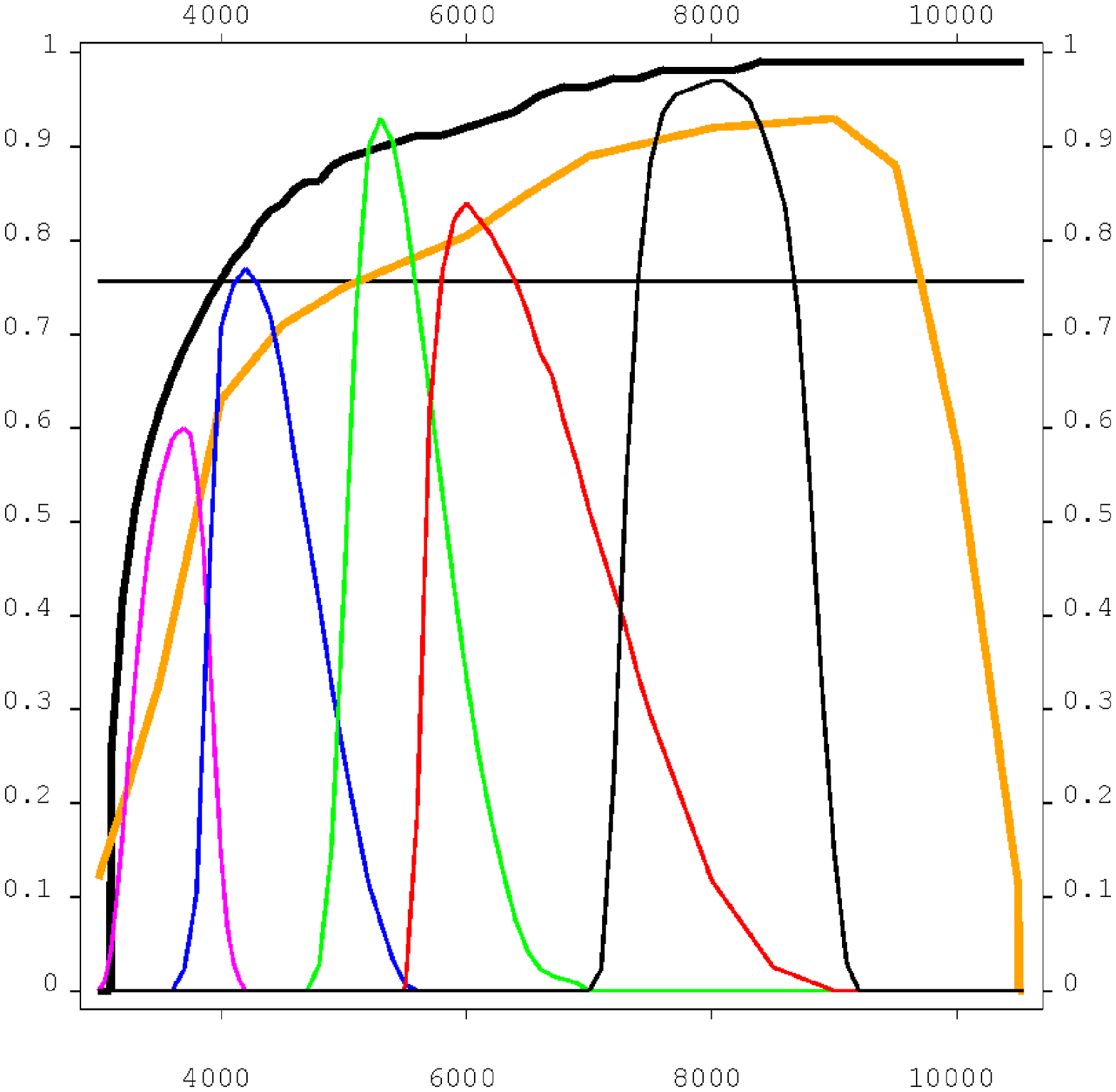,width=14.cm,height=8.cm}}
\mbox{\epsfig{file=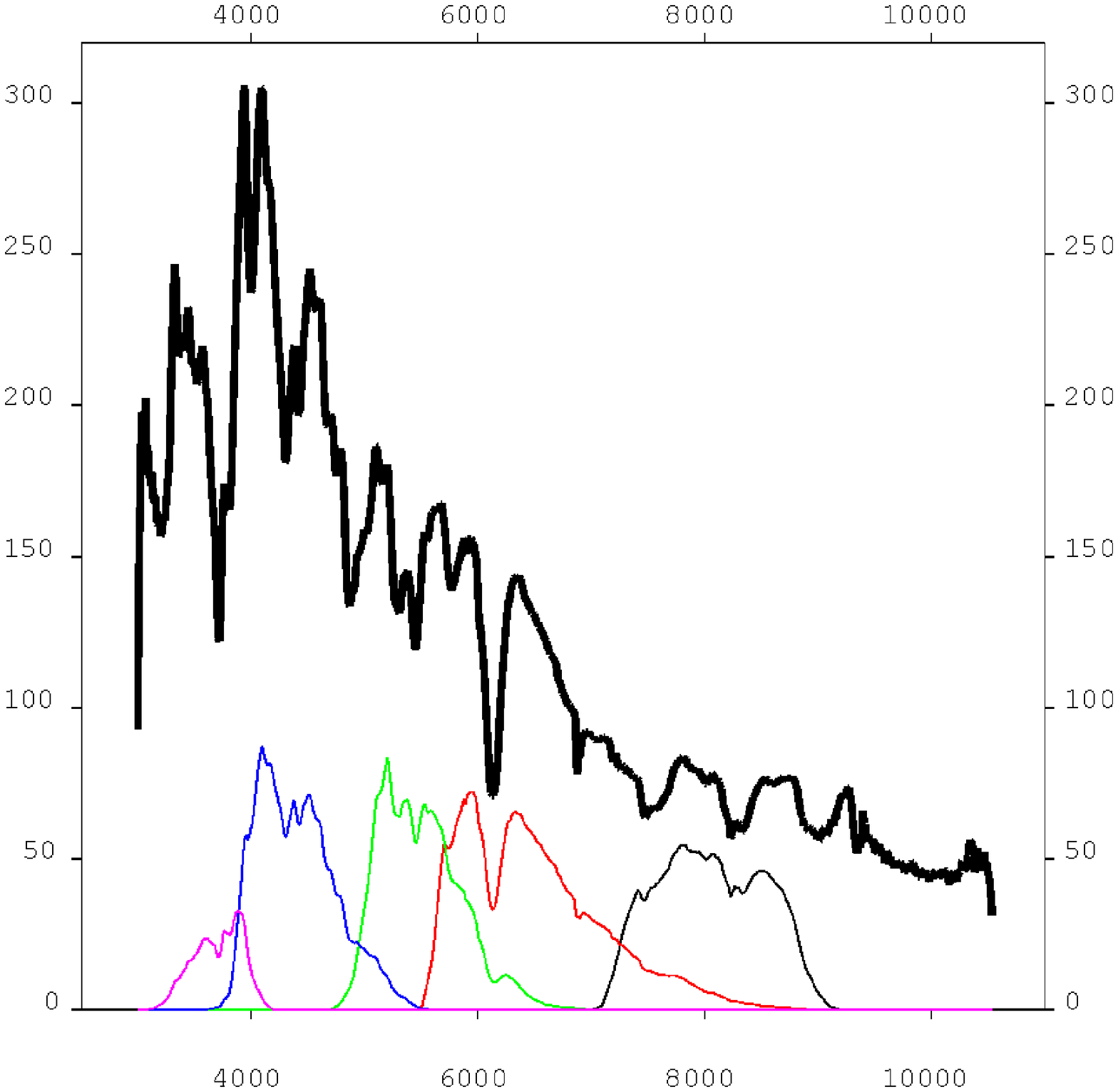,width=14.cm,height=8.cm}}
\caption[]{\it Upper panel: transmission of the {\bf UBVRI} filters (coloured lines),
the atmosphere (black thick), the optics (black standard), and quantum
efficiency of the CCD (orange thick) as a function of the wavelength.\\
Lower panel: measured SN1994D spectrum in $photons/m^2/s/$\AA,
3 days after maximum (black thick line),
residual spectras after atmospheric, instrumental and filter transmissions, 
and conversion in the CCD in $\gamma e^-/m^2/s/$\AA\ (coloured lines).
\\
\label{figSPEC}}
\end{center}
\end{figure}

By definition, $\Phi_w(M_w=0)$ depends on the spectrum of the measured
point-like object. For SNIa's,
we estimate the number of photoelectrons produced in the detector
in each of the {\bf UBVRI} passbands by using the following ingredients:
\begin{itemize}
\item
The photon spectra and the magnitudes
of two typical SNIa's:\\
SN1990N, 7 days before and 7 days after maximum\cite{sn1990n},
and SN1994D, 3 days after maximum\cite{sn1994d}.
\item
The atmospheric transmission curve taken at airmass=1.
\item
The Bessel {\bf UBVRI} theoretical transmission curves.
\item
The LBNL red-enhanced CCD spectral response.
\item
The transmission of the MARLY F/8 optics, which has been assumed to be
uniform in wavelength, equal to 0.87 per reflector,
i.e. $0.87^2=0.76$.
\end{itemize}
Fig. \ref{figSPEC} shows the various transmission curves and the
SN1994D spectrum.
From this input, we find that a SNIa with $M_{\bf B}\simeq M_{\bf V}=0$, and a
spectrum typical of the one emitted within +/-7 days from maximum, 
should produce a flux of photoelectrons in the detector given
by the upper part of Table 1.
\begin{table}
\begin{center}
\caption[]{\it The expected total 
$\gamma e^-$ fluxes produced by the LBNL red-enhanced CCD at Airmass=1
for a SNIa at maximum in {\bf B}. The collecting area of the MARLY is
$S=0.57\ m^2$.\\}
\begin{tabular}{|c|c|c|c|c|c|c|}
\hline
\multicolumn{7}{|c|}{}\\
\multicolumn{7}{|c|}{$\Phi_w({\bf B}=0)=$Flux ($\gamma e^-/m^2/s$) for SNIa with {\bf $M_B=0$}}\\
\multicolumn{7}{|c|}{}\\
\hline
\multicolumn{2}{|c|}{Filter w} &{\bf U} &{\bf B} &{\bf V} &{\bf R} &{\bf I}\\
\hline
\multicolumn{2}{|c|}{Magnitude} & -0.58 & 0. & 0.09 & 0.03 & 0.29\\
\multicolumn{2}{|c|}{$\Phi_w({\bf B}=0)\ (10^9\ \gamma e^-/m^2/s)$} & $0.9$ & $4.$ & $3.9$ & $5.2$ & $3.7$ \\
\hline
\multicolumn{7}{|c|}{}\\
\multicolumn{7}{|c|}{$\Phi_w(M_w=0)=$Flux ($\gamma e^-/m^2/s$) for SNIa with {\bf $M_w=0$}}\\
\multicolumn{7}{|c|}{}\\
\hline
\multicolumn{2}{|c|}{Magnitude} & 0. & 0. & 0. & 0. & 0. \\
\multicolumn{2}{|c|}{$\Phi_w(M_w=0)\ (10^9\ \gamma e^-/m^2/s)$} & $0.54$ & $4.$ & $4.2$ & $5.3$ & $4.8$ \\
\hline
\hline
\multicolumn{7}{|c|}{}\\
\multicolumn{7}{|c|}{Expected total flux for the MARLY ($\gamma e^-/s$)}\\
\multicolumn{7}{|c|}{$F^{SN}_w=S\times\Phi_w(B=0)\times 10^{-M_B/2.5} $}\\
\multicolumn{7}{|c|}{}\\
\hline
redshift & at max &{\bf U} &{\bf B} &{\bf V} &{\bf R} &{\bf I}\\
\hline
0.01 & Magnitude    & 13.3 & 13.9 & 14.0 & 13.9 & 14.2 \\
     & $F^{SN}$ & 1440 & 6280 & 6120 & 8160 & 5810 \\
\hline
0.05 & Magnitude    & 16.8 & 17.4 & 17.5 & 17.4 &  17.7 \\
     & $F^{SN}$ & 58   & 250  &  244 &  325 &  231 \\
\hline
0.1  & Magnitude    & 18.3 & 18.9 & 19.0 & 18.9 & 19.2 \\
     & $F^{SN}$    & 14   & 63   & 61   & 82   & 58   \\
\hline
\end{tabular}
\end{center}
\label{tab1}
\end{table}
%
Taking into account the measured colour indices of SN1994D,
we get a correspondance between the zero $M_w$ magnitudes
and the photoelectron fluxes in {\bf UBVRI} (lower part of Table 1)
\footnote{These estimates are compatible with the {\it observed} signal on
SNIa images taken with the Danish telescope. But only a rough check
could be done because of differences in the filter transmissions
and CCD efficiency.}.

\subsection{SN photometry simulation}
We have computed the total time spent in the photometric follow-up 
using a simple SN light curve simulation. 
We suppose all SNe to lie at $z=0.05$. 
The whole calculation is done for airmass=1.
The SN discovery age is supposed to range between 5 and 10 days 
before the time of {\bf B} maximum with a flat distribution.
The lunar phase is also randomly chosen.
\subsubsection{SN environnement - Expected photoelectron background}
The background surface magnitude due to both the sky brightness
and the host galaxy surface brightness in the w passband is given by:
$$\mu_w = -2.5 log_{10}\left(10^{-\mu_w^{gal}/2.5}+10^{-\mu_w^{sky}/2.5}\right)$$
When the background flux is negligible (i.e. $\mu_w$ is large, or if the
SN is bright), the last factor in the expression of $T_{exp}$
is 1, and the formula can be simplified.
\begin{itemize}
\item{\bf Night sky brightness $\mu_w^{sky}$}\\
The sky brightness depends on the Moon phase and also on the atmospheric
conditions. We adopt the average values used for the preparation of
the ESO proposals (Table 2, upper part).
\begin{table}
\begin{center}
\caption[]{\it Sky brightness in magnitudes for different passbands, as a
function of the number of days from the New Moon;
range and average of the host galaxy brightness.
All magnitudes are the ones at the top of the atmosphere.\\}

\begin{tabular}{|c|c|c|c|c|c|}
\hline
Days from &\multicolumn{5}{|c|}{Sky brightness}\\
New Moon  & {\bf U} &{\bf B} &{\bf V} &{\bf R} &{\bf I}\\
\hline
0  & 22.0 & 22.7 & 21.8 & 20.9 & 19.9 \\
3  & 21.5 & 22.4 & 21.7 & 20.8 & 19.9 \\
7  & 19.9 & 21.6 & 21.4 & 20.6 & 19.7 \\
10 & 18.5 & 20.7 & 20.7 & 20.3 & 19.5 \\
14 & 17.0 & 19.5 & 20.0 & 19.9 & 19.2 \\
\hline
\hline
\multicolumn{6}{|c|}{host galaxy brightness}\\
\hline
min.  & 22. & 22. & 21.5 & 21. & 20.5 \\
{\bf average} & {\bf 21.} & {\bf 21.} & {\bf 20.5} & {\bf 20.} & {\bf 19.5} \\
max.  & 19. & 19. & 18.5 & 18. & 17.5 \\
\hline
\end{tabular}
\end{center}
\label{tab2}
\end{table}
\item{\bf Host galaxy brightness $\mu_w^{gal}$}\\
The background light from the host galaxy depends on the
position of the SN. If the SN is in the disk, far from the bulge,
then the average surface brightness of the galaxy in {\bf B} is
$\mu^{gal}_{\bf B} = 22.4$ \cite{surface}.
If the SN is located in the bulge, the average is $\mu^{gal}_{\bf B} \sim 20$.
The SNIa's referred to in the EROS paper\cite{tauxSN}
have a local host surface brightness
ranging between $\mu^{gal}_{\bf B} = 19$ and $\mu^{gal}_{\bf B} = 22$,
and more than 80\% have $\mu^{gal}_{\bf B} > 20$ (see Fig. \ref{figHOST}).
Following \cite{surface} we adopt $<\mu^{gal}_{\bf U}-\mu^{gal}_{\bf B}>\sim 0$ and
$<\mu^{gal}_{\bf B}-\mu^{gal}_{\bf R}>\sim 1$ as typical values, and we extrapolate
for V and I.
\begin{figure}
\begin{center}
{\includegraphics[width=.8\textwidth]{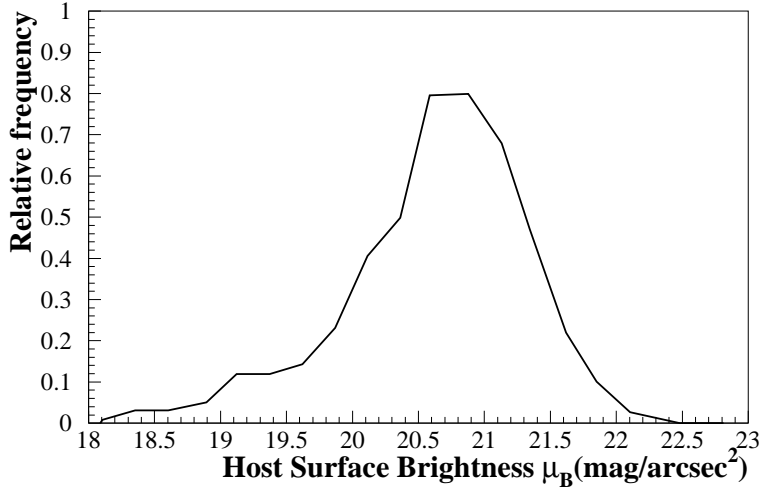}}
\caption{\it\label{figHOST} 
Distribution of the host galaxy {\bf B}-magnitude in the EROS SNIa
discovery sample.}
\end{center}
\end{figure}
In the following, we
will assume that the SNIa's that we will follow-up
will belong to such a sample.
\end{itemize}
\subsubsection{SN light curve simulation}
Light curves of SNIa's at $z=0.05$ are generated according to
the observed frequency of various templates described in \cite{regnault}.
\begin{figure}
\begin{center}
{\includegraphics[width=\textwidth,clip=true]{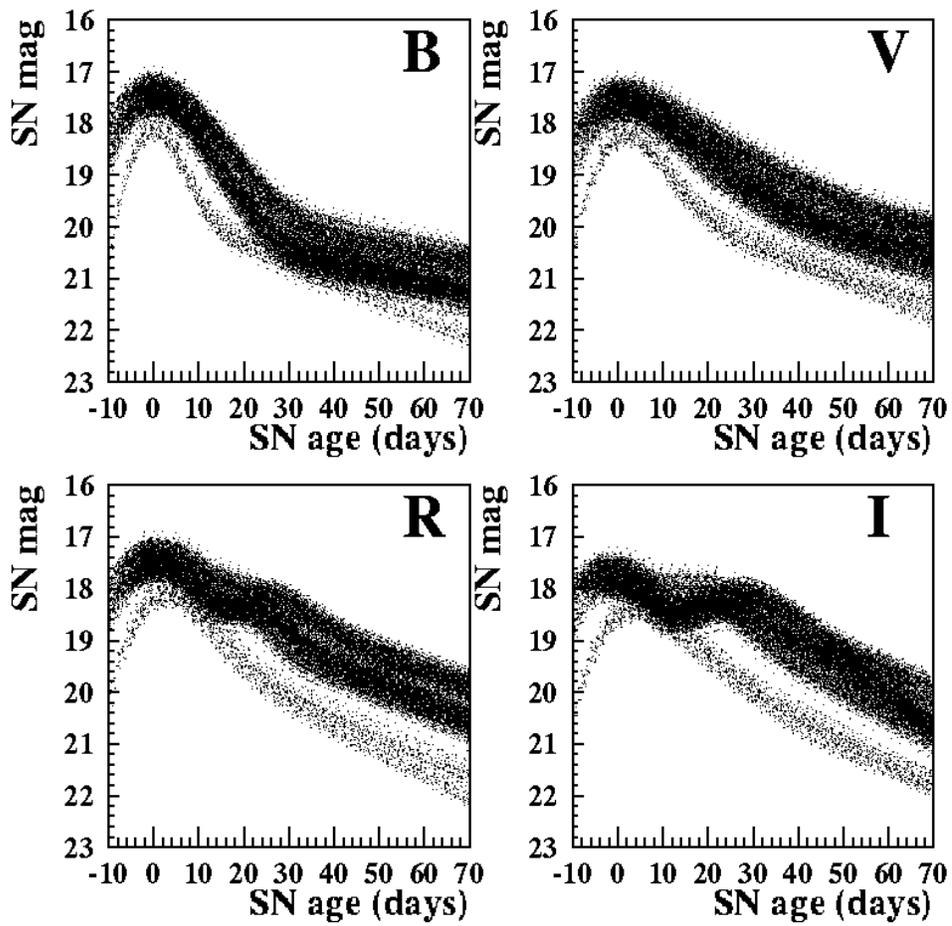}}
\caption{\it\label{figCL} 
SN Light curves (in mag) obtained in the simulation.}
\end{center}
\end{figure}

SNIa light curves show some diversity.  
We use {\bf B},{\bf V},{\bf R} and {\bf I} templates obtained in \cite{regnault} to describe 
the different luminosity evolutions with time. We used those templates  
from reference \cite{regnault} for which early data is available
\footnote{SN 91T, 91bg, 92A, 94D, 95D, 96X, 98bu, 98de}.
The absolute peak magnitude attributed to each template is based on
the standardisation relation determined in \cite{regnault}. 

\begin{figure}
\begin{center}
{\includegraphics[width=.8\textwidth]{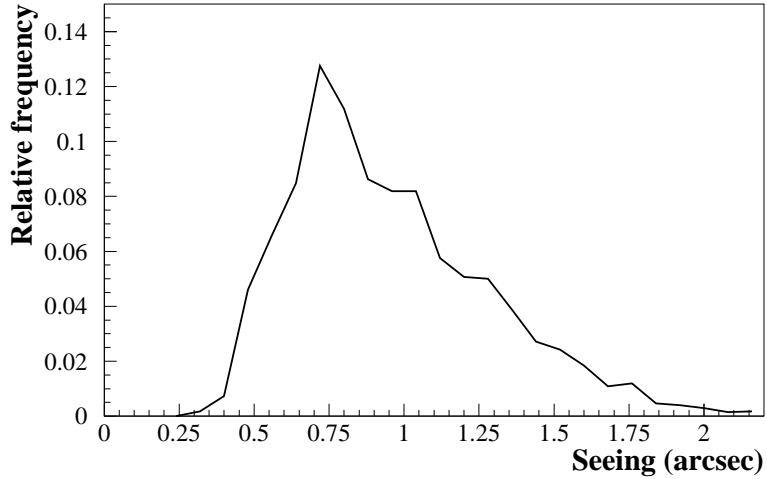}}
\caption{\it\label{figSEEING}
Distribution of the seeing at La Silla (in arcsec).}
\end{center}
\end{figure}

\begin{figure}
\begin{center}
{\includegraphics[width=\textwidth]{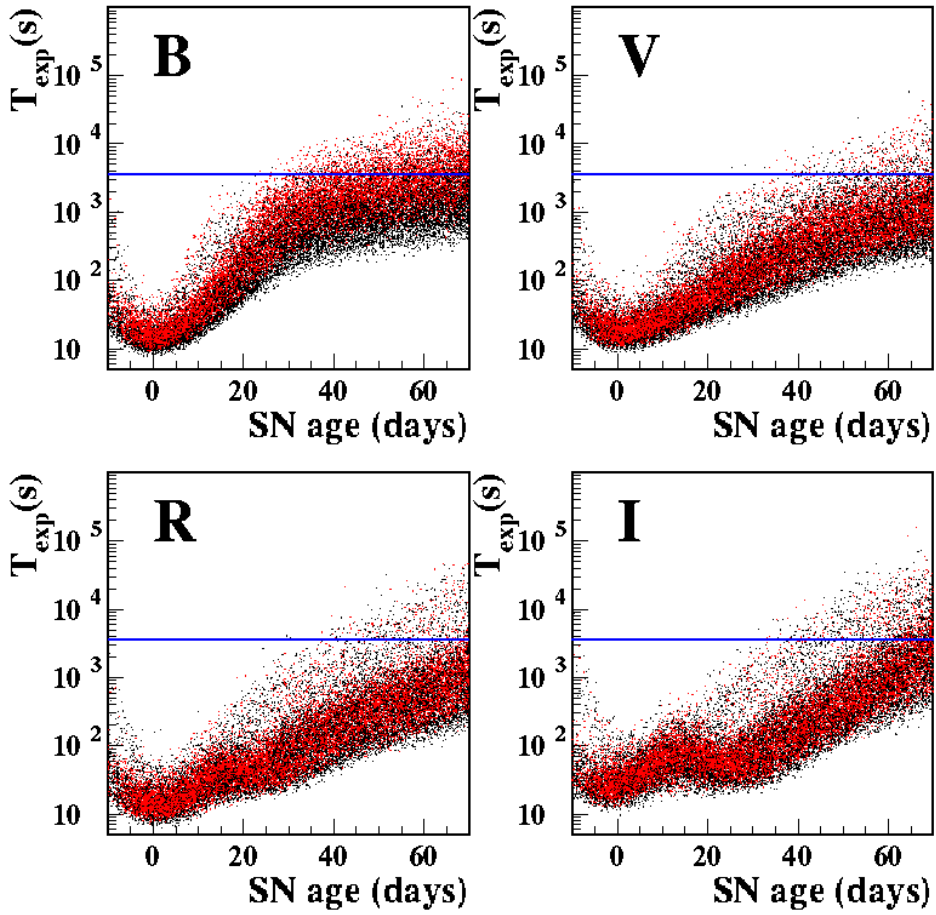}}
\caption{\it\label{figTEXP}
$Log_{10}(T_{exp})$ needed to reach 2\% photometric precision versus the SN age
(in days).
Exposures taken around full moon ($\pm 4$ days) are shown in red. 
The horizontal lines illustrates the 1 hour cutoff used in this study.
}
\end{center}
\end{figure}

The program randomly generates a template according to the
frequency distribution which came out of in the best fit
analysis of \cite{regnault}. 
The SN maximal luminosity is  smeared 
using a normal law with 0.1 mag dispersion (in each band).
For each SN, a host surface brightness is generated according to 
the distribution shown in Fig. \ref{figHOST}.
Once a discovery age, lunar phase, host brightness 
and SN subtype have been determined, 
the program follows the SN luminosity evolution for 70 days after B maximum. 
Photometric measurements are assumed to take place every other night,
and a 70\% good weather is assumed.  
In a bad weather case, the new measurement is attempted again the next day. 
1000 light curves obtained this way are shown in figure \ref{figCL}.
We associate a random seeing to each measurement,
using the true seeing distribution of La Silla
(see Fig. \ref{figSEEING});
an instrumental contribution
of $0.8''$, as measured on the MARLY optical axis,
is quadratically added to this seeing.
The exposure time necessary to reach a 2\% photometric 
precision is then computed for each measurement.
These times are shown in figure \ref{figTEXP}.
These exposure calculations are done for the MARLY telescope.
The scaling to be performed in order to get results for another
telescope (at the same site) or another resolution is straightforward.
\begin{figure}
\begin{center}
{\includegraphics[width=\textwidth]{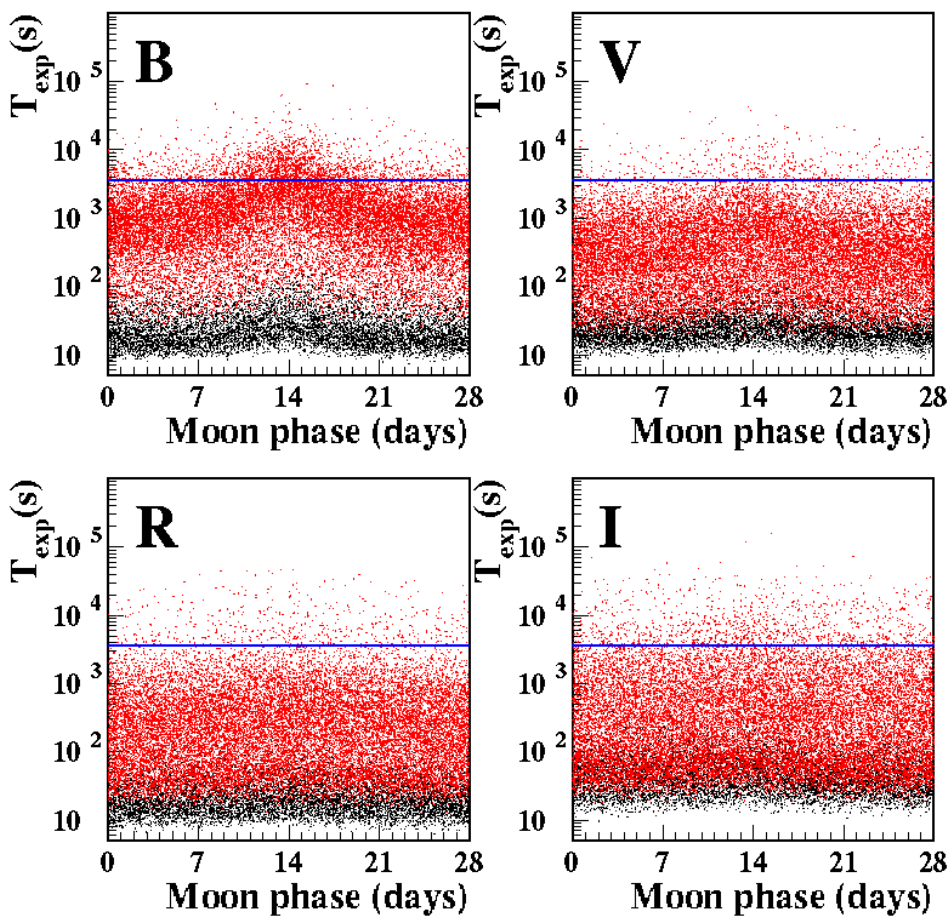}}
\caption{\it\label{figLUNE} 
$Log_{10}(T_{exp})$ needed to reach 2\% photometric precision
versus the Moon phase (in days).
The red dots 
correspond to $SN\ age > 15\ days$. For these late measurements,
the exposure times are much longer and strongly depend on the Moon phase.
The horizontal lines illustrates the 1 hour cutoff used in this study.}
\end{center}
\end{figure}
\begin{figure}
\begin{center}
{\includegraphics[width=\textwidth]{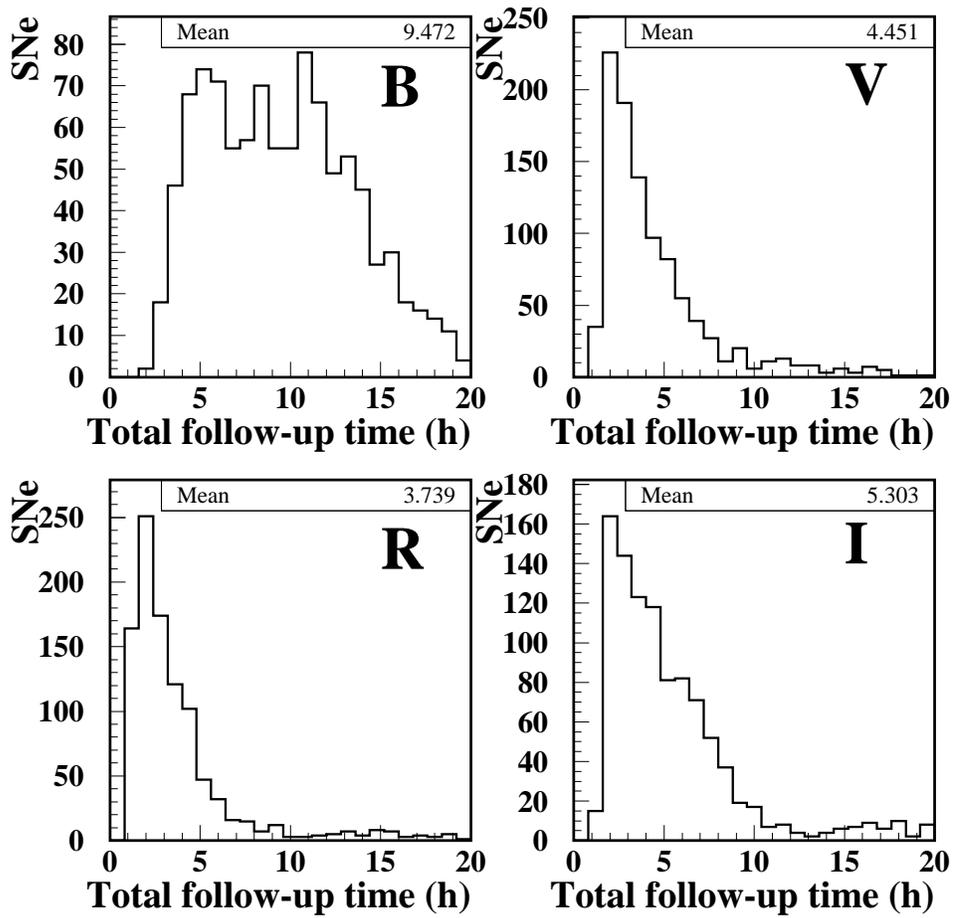}}
\caption{\it\label{figTT}  
Distribution of the total
follow-up time per simulated SNIa, in hours.
Each entry corresponds to the sum of the exposure times of 
about 40 measurements with 2\% photometric precision.} 
\end{center} 
\end{figure}
\subsection{The ``cost'' of a SNIa}
Fig. \ref{figTEXP} and \ref{figLUNE} give the distribution of exposure times as a
function of SN age and Moon phase respectively.
Long exposure times occur when the SNIa fades, typically from
15 days after maximum (red dots in Fig. \ref{figLUNE}). The exposure
then strongly depends on the sky brightness in {\bf B} and {\bf V}.

\begin{figure}[t]
\begin{center}
{\includegraphics[width=.8\textwidth]{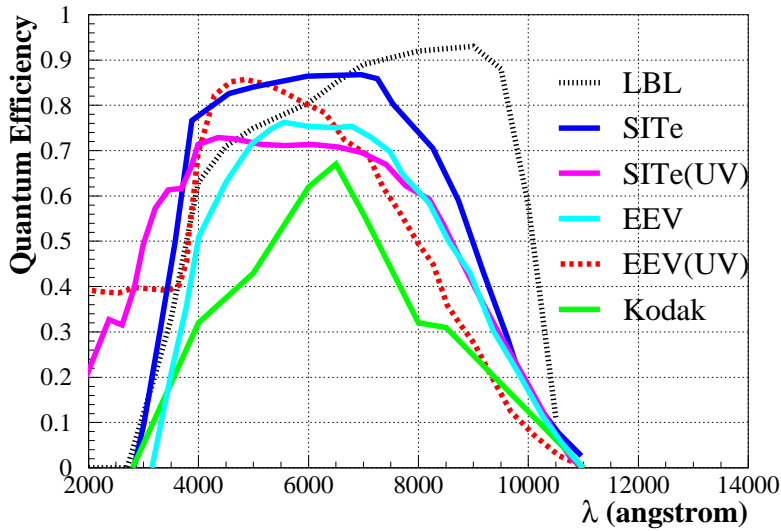}}
\caption{\it\label{figCCD} CCD efficiencies used in the calculation.}
\end{center}
\end{figure}

We estimate the telescope time needed per SNIa by adding the
exposure times of all the measurements (about 40 measurements per SNIa and
per passband). An overhead of 60 s is added to
each exposure. For long exposures, which will be fractionned,
we further add 60 s of overhead to every sequence of 20 minutes
of observation.
We choose to truncate the exposures at 3600 seconds. Such a truncation
affects 7\% of the measurements and takes place only in the late part of the SN
light curves ($age > 30\ days$). The
photometric resolution is then slightly downgraded (from 2\% to 4\% in the
worst case).
Fig. \ref{figTT} shows the histogram of the total telescope time
for each passband, for our sample of 1000 simulated SNIa's.
The outcome of this study is that the follow-up of a SNIa, involving about
40 measurements regularly spaced within a 80 day period,
needs an average telescope time of:
\begin{itemize}
\item
9.5 hours in {\bf B},
\item
4.5 hours in {\bf V},
\item
3.8 hours in {\bf R},
\item
and 5.3 hours in {\bf I},
\end{itemize}
Table 3 gives the results of the same calculations done for other
CCD efficiencies (see Fig. \ref{figCCD}).
\begin{table}[h]
\begin{center}
\caption{\it Total follow-up time (hours) for 4 CCD efficiencies, 
in the {\bf BVRI} bands}.
\begin{tabular}{c|cccc|c}
CCD  & {\bf B} & {\bf V} & {\bf R} & {\bf I} & Total\\
\hline
\hline
Berkeley &  9.5h & 4.5h & 3.8h &  5.3h & 23.1 \\
SITe     &  8.6h & 4.2h & 3.7h &  6.2h & 22.7 \\
EEV      & 10.6h & 4.6h & 4.2h &  7.0h & 26.4 \\
Kodak    & 14.1h & 6.1h & 4.9h &  9.4h & 34.5 \\
\end{tabular}
\end{center}
\end{table}
\subsection{Strategy}
In order to minimize the cost of the measurements taken during the
late section of a SNIa (when faintest)
we could consider some improvement to the very simple strategy described
above:
\begin{itemize}
\item
As long as the SNIa (at $z=0.05$) is close to its maximum ($\sim 17.5$), one
can neglect the sky brightness in {\bf BVRI}
and the host galaxy luminosity in a first
approximation. This is no longer the case when the SN fades.
Then, when the Moon is bright, we should
favour data taking on the brightest SNIa's, and devote the dark fraction
of each night to the observation of the faintest ones.
\item
As the luminosity variations of ``old'' SNIa's are slow,
relaxing the sampling rate should have less impact than around the maximum.
\item
Lastly, we could downgrade the photometric precision when the estimated
exposure time is long. The impact of
such a choice should be studied.
\end{itemize}
\subsection{An additional dedicated camera for {\bf U}}
The {\bf U} band is of special interest, because the detailed study of the
light curves of the nearby SNIa's will be useful for the
{\bf I} band study of high redshift SNIa's ($z>1$).

\begin{figure}\begin{center}
\includegraphics[width=\textwidth,clip=true]{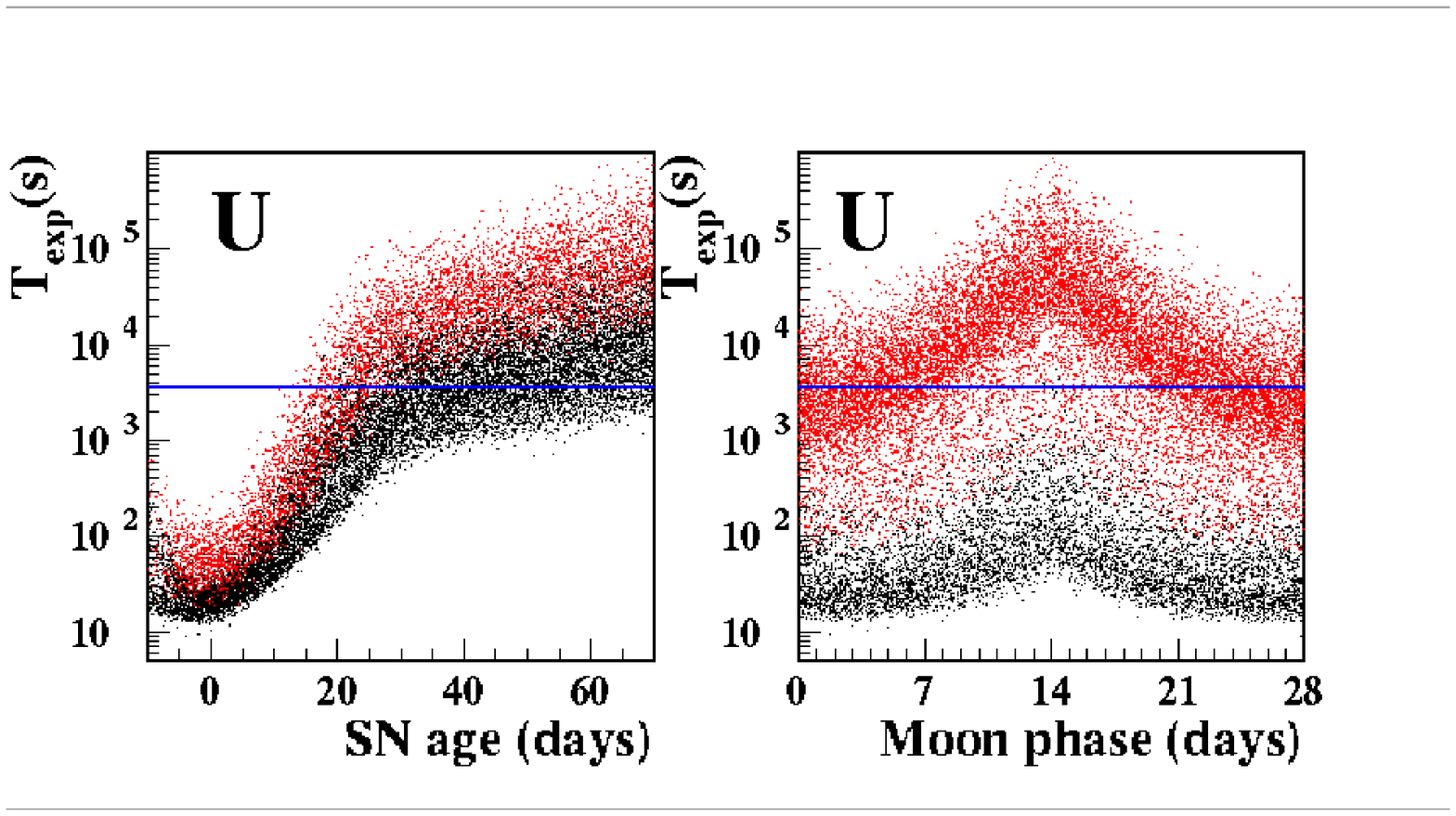}
\caption{\it\label{figU} Exposures needed to reach a 2\% photometric precision in {\bf U} 
versus the SN age (left) and Moon phase (right) with the Berkeley CCD. The colors 
are drawn as for figures \ref{figTEXP} and \ref{figLUNE}. 
The horizontal lines correspond  to 1 hour.}
\end{center}\end{figure}

For this passband, only two SNIa templates are available. The first order
calculation performed in the first version of this note (May 2001) showed
that more than 60 hours were needed to follow-up a SNIa in {\bf U} band,
with our simple strategy. 
Figure \ref{figU} shows the individual exposures in {\bf U}, computed as 
for the other filters for the Berkeley CCD, as a function of the SN 
age and lunar phase. 

\begin{figure}[t]
\begin{center}
\includegraphics[width=.8\textwidth]{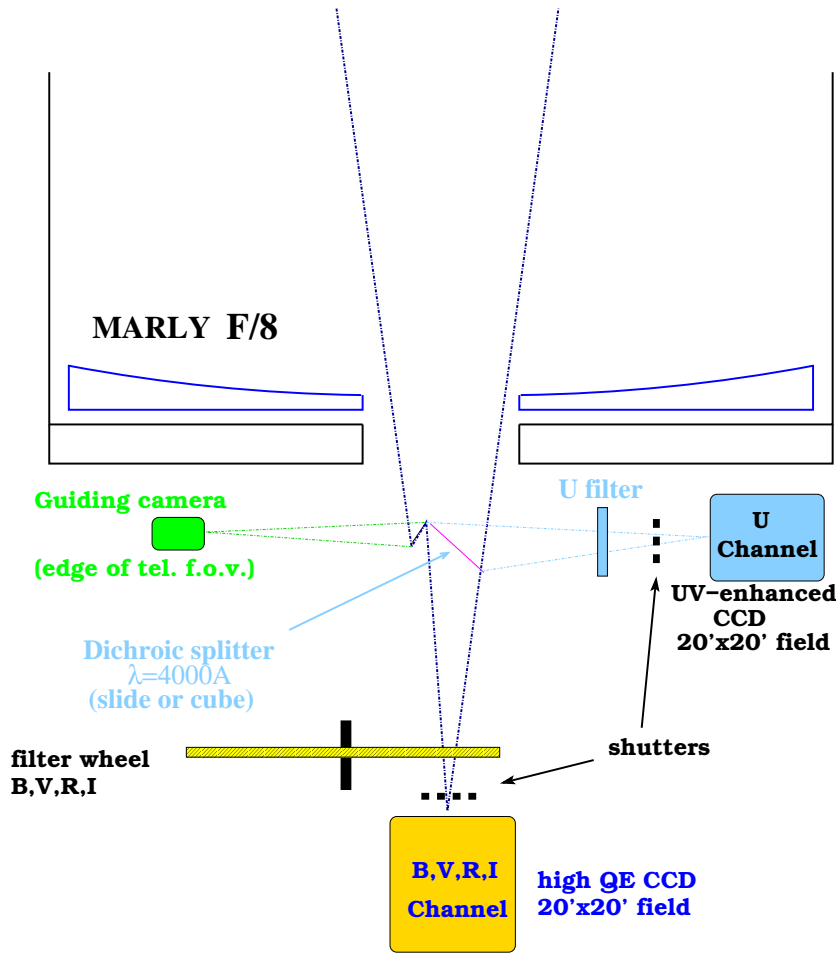}
\caption{\it\label{fig7UP} A possible setup for a photometer integrating 
in {\bf U} 
and in {\bf BVRI} in parallel}
\end{center}
\end{figure}

Since unrealistically long exposures are needed for the {\bf U} band,
we have considered the possibility to take images in parallel in
the {\bf U} band and in {\bf BVRI} with a dichroic beam splitter. 
A possible setup to achieve this 
is displayed on figure \ref{fig7UP}. Preliminary discussions with 
optics engineers from Observatoire de Marseille confirmed the 
technical feasibility of this setup. 

\begin{figure}
\includegraphics[width=\textwidth]{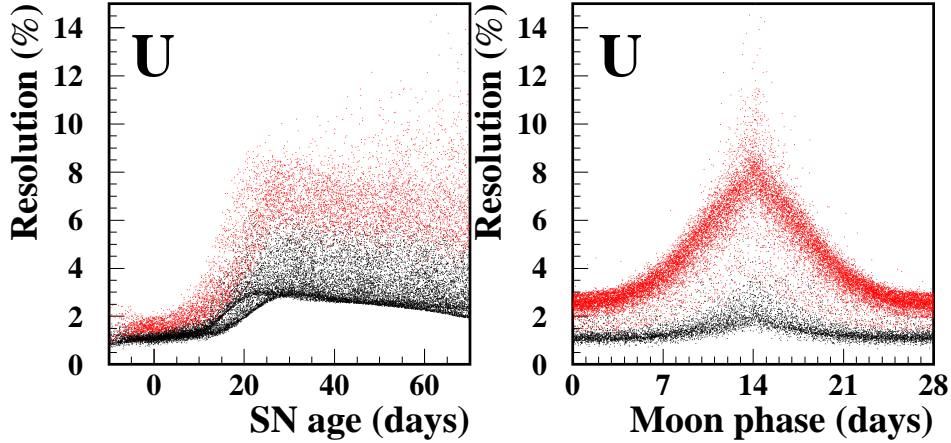}
\caption{\it\label{figUEFF} Photometric resolutions achieved 
when exposing in {\bf U} with  a 
SITe UV-enhanced device, in parallel with {\bf BVRI} and the LBL CCD, 
as a function of SN age and Moon phase. Red points indicate 
measurements scheduled near full Moon (left) or at 
a SN older than 15 days (right).}
\end{figure}

In this setup the CCD detectors can be separately 
optimized for the {\bf U}  and the {\bf BVRI} bands.
The data taking could proceed as follows. Consecutive exposures in {\bf BVRI} 
are chosen so as to ensure a 2\% photometric resolution. 
The exposure in {\bf U} lasts until the end of this cycle. 
To evaluate the viability of this scheme, 
we have estimated the resulting photometric precision in {\bf U}. 
We have assumed that the UV-enhanced SITe CCD will be used 
for this channel (see figure \ref{figCCD}). 
Figure \ref{figUEFF} shows these individual resolutions 
as a function of the SN age and lunar phase.  

\begin{figure}[t]
\begin{center}
\includegraphics[width=.8\textwidth]{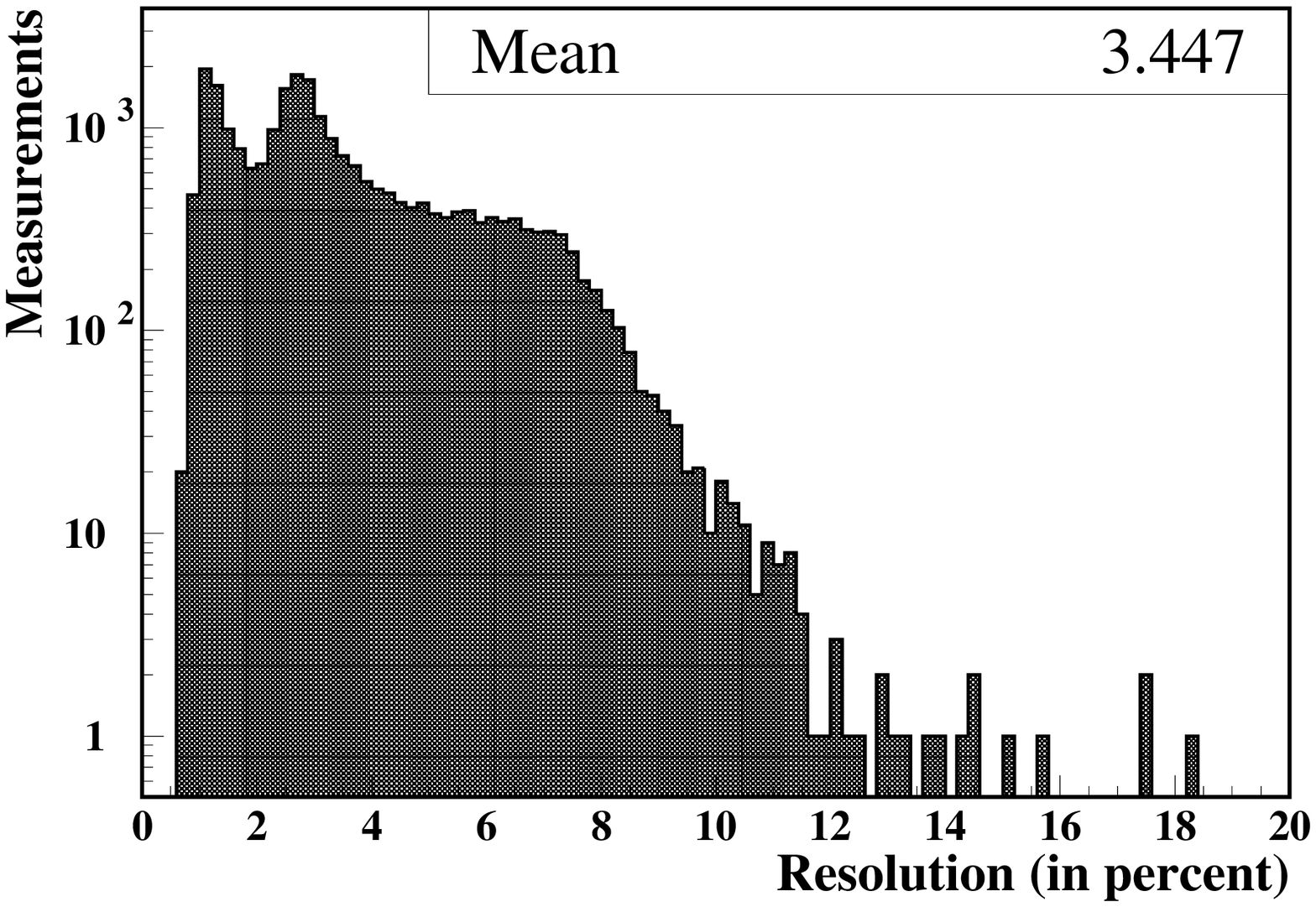}
\caption{\it\label{figURES} Photometric resolutions achieved 
when exposing in {\bf U} with  a 
SITe UV-enhanced device in parallel with {\bf BVRI} and the LBL CCD. 
This corresponds to a strategy aiming at a 2\% resolution in {\bf BVRI}.
}
\end{center}
\end{figure}
The moon has clearly a strong degrading effect (near full moon) ; 
late epochs (after day 25) are also attributed a worse resolution.

Figure \ref{figURES} shows the resolutions  achieved with
 the naive strategy discussed here.
Almost all measurements have a 
resolution better than $\approx 10\% $. The average resolution is $\approx 3.5\%$. 
A more subtle observing strategy could be sought to optimize 
the achieved resolutions versus the observing time. 
For example, one could decide to sometimes impose the 
{\bf BVRI} exposures to reach a fixed, better resolution in {\bf U}.


The numbers obtained here in the {\bf U} passband are 
somewhat uncertain because
of 
the large fluctuations of the atmospheric absorption. 
However, since exposures in {\bf U} and other bands are simultaneous, 
color dependent absorption corrections due 
to variations in atmospheric conditions should be controlled.
Monitoring these absorption corrections will be easier with a 
larger CCD field ;  $2k\times 2k$ chips are at least necessary \footnote{In the $F/8$ 
configuration of the MARLY, and with $15\mu$ pixels, 
this corresponds  to $13'\times 13'$ on the sky}.

Finally, it is worth to mention that 
a factor of 2 in the photo-electron flux in {\bf U} could be gained 
by removing the {\bf U} filter. The effective passband (limited by 
the atmospheric absorption cutoff and the dichroic) would then 
be slightly modulated by the atmospheric conditions. 
This solution, still under study, might be considered only if the absorption 
is proved to be precisely monitored.


\subsection{Maximum number of SNs that can be monitored}

 Data will be taken
sequentially with the {\bf BVRI} 
filters. The total telescope time
per SNIa will thus 
be: 9.5+4.5+3.8+5.3=23.1 hours, spread over a period of 80 days.
The average duration of the astronomical night at La Silla is 9 hours,
and the average clear sky fraction is 70\%. Then the average available
time per period of 80 days is 80x9x0.7=504 hours. Up to 22 SNIa's
can be completely measured during this period of 80 days.
Therefore the rate of 100 SNIa's per year completely measured
in the {\bf UBVRI} passbands is the capacity of the MARLY 
with the setup outlined in figure \ref{fig7UP}. 
This rate could be made larger with a more subtle strategy as indicated
above with a limited loss in  sampling or in photometric
precision. By using the tables and figures provided in this note,
the reader can estimate the performance that would correspond
to another strategy.

\subsection{Miscellaneous}
One should keep in mind that
we have considered SNIa's with no reddening in our calculations. In the case
of reddening, the estimates in the blue passbands may be significantly
affected.

For the {\bf U} band, airmass is a critical issue since the atmospheric
absorption is a dominant factor\footnote{$\Delta{\bf U}=0.44/airmass$
and $\Delta{\bf B}=0.23/airmass$ at airmass=1,
for a SNIa spectrum and the CCD spectral efficiency adopted here.}.
The signal from the SN and the host
galaxy would be lowered by a factor 1.22 if the airmass were 1.5 instead
of 1, and it is clear that the full Moon would strongly disturb the
observations in this passband.

Due to the presence of a dichroic splitter, the effective {\bf B} band 
could be slightly narrower 
than the standard one. 

From figure \ref{figSPEC} one may see that this setup is also 
sensitive in the {\bf Z} band which could be added to our observation scheme 
in addition with {\bf BVRI}. Nearby SNIa have not yet been thoroughly 
studied in this band which could be helpful to 
disentangle the SN reddening by the host galaxy's dust and the intrinsic 
SN colour dispersion. 
Finally, let us recall that 
the proposed configuration could be optimized to other 
photometric systems (e.g. $ugriz$).

\subsection{Necessary upgrades and schedule}
To use the MARLY as a photometer, according to our preliminary design, 
the following technical changes
must be done:
\begin{itemize}

\item
The focal reducer has to be removed, and the F/8 optics has to be
re-installed. A dichroic beam splitter at a wavelength near 4000 \AA\ 
should be installed.
Moreover, the guiding system has to be adapted.

\item two CCD cameras have to be built :
\begin{itemize}
\item
A single CCD camera with a high performance CCD 
for {\bf BVRI},

\item
A single CCD camera with a UV-enhanced CCD  
for {\bf U}.
\end{itemize}
\end{itemize}

\noindent
The EROS project will stop taking data at the end of december 2002. 
Designing and building the instrument can be largely 
decoupled from MARLY running and thus could be performed 
in France. Installing and testing the system on site 
should not take longer than one month, as in the EROS2 case. 
Provided a more general framework is found for this project, 
a startup as early as spring 2003 could take place. 

\subsection{Expected performance}
Around {\em 100 SNIa's} per year at $z\sim 0.05$
could be fully followed-up in {\bf BVRI} with a photometric precision $\delta =2\% $, 
and in the same time in {\bf U} with a 3.5\% photometric precision on average.
The exposure time scales with 
$\delta^{-2}$. If the overhead time is small compared to the
exposure time (if $\delta<2\% $), then the number of SNIa's that can be
monitored is $\sim 25\times\delta(\%)^2$ per year.
If $\delta>2\%$, then the overhead time is not negligible and we
have to use the following expression
$$T_{telescope} \sim 40\times(<T_{exp}>+T_{overhead})$$
to correctly estimate the corresponding telescope time and the rate
of monitored SNIa's.
We thus find that around
{\em 300 SNIa's} per year at $z\sim 0.05$ could be fully followed-up
with a $\delta=4\%$ photometric precision in {\bf BVRI}.

\section{Conclusion}

Other possible uses (as a SN discoverer or with an integral field spectrometer) 
of the 1 meter dedicated MARLY telescope, based at
La Silla, have also been evaluated in a more complete studiy 
available on request. 

Refurbished in order to simultaneously perform {\bf BVRI} and 
{\bf U} photometry, the MARLY
could be used for the complete follow-up (40 measurements spanning 80 days) 
 of at least 100 SNIa's per year 
at $z\sim0.05$,
with a photometric precision of 2\% in {\bf BVRI} 
and 3.5\% on average in {\bf U}.
If the required photometric precision is only 4\%, then
about 300 SNIa's can be followed-up each year. 


A project dedicated to the nearby SN study should 
integrate dedicated discovery and follow-up means, 
the latter providing spectrographic and/or photometric capabilities. 
With the equipment proposed in this note,  
the MARLY telescope could nicely complement such a project.
We will would be happy to collaborate on any such project.



\newpage
\section*{Annex A\\
Expression of the optimal photometric precision
accessible with aperture\\
photometry}
\begin{figure}
\begin{center}
\mbox{\epsfig{file=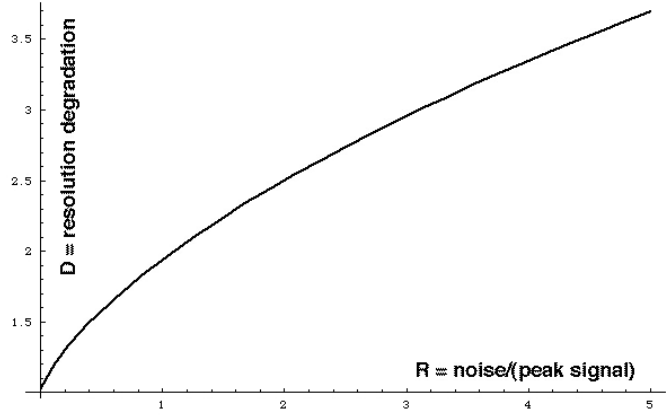,width=9.5cm,angle=0}}
\caption[]{\it The degradation factor D of the resolution as a function
of the ratio R of the noise per pixel to the maximum signal
per pixel.
\\
\label{fig1}}
\end{center}
\end{figure}

Let
\begin{itemize}
\item
$T_{exp}$ be the exposure time (in s),
\item
$F^{SN}$ be the integrated photoelectron flux from the SN
in the telescope in photo-electrons per second ($\gamma e^-/s$),
\item
$b$ be the photoelectron flux per pixel originating in
the atmospheric emission and the host galaxy (in $\gamma e^-/s$)
\item
$p$ be the pixel size.
\end{itemize}
Assuming that the PSF is a gaussian of standard deviation
$\sigma=seeing/2.36$:
$$PSF(r)=\frac{F^{SN}\times T_{exp}}{2\pi\sigma^2}e^{-r^2/2\sigma^2},$$
and that {\it the pixel size is small with respect to the seeing},
the time integrated photoelectron signal in a disk of radius r
is given by
$$S(r)=F^{SN}\times T_{exp}\times (1-e^{-r^2/2\sigma^2}).$$
During the same exposure time, the number of photoelectrons due to
the atmosphere and the host galaxy in the disk of
radius r is given by
$$B(r)=b\times T_{exp}\times \pi \left[\frac{r}{p}\right]^2.$$
Assuming that we measure separately $<B(r)>$ with a negligible
uncertainty, then
the fluctuation of the sum $S(r)+B(r)-<B(r)>$
is given by
$$\sigma(r)^2=S(r)+B(r)+\sigma_e^2\times\pi \left[\frac{r}{p}\right]^2,$$
where $\sigma_e$ is the readout noise per pixel.
The photometric resolution is then given by the ratio
$\delta=\frac{\sigma(r)}{S(r)}$.
The resolution $\delta$ has its smallest value when $\frac{d\delta}{dr}=0$.
Expressing this condition leads to the calculation of the optimal
value of r, and the expression of the best resolution follows:
$$\delta=[F^{SN}\times T_{exp}]^{-\frac{1}{2}}\times D,$$
where D is a ``degradation'' factor of the resolution, depending
only on the noise to signal ratio in the {\it pixel where the signal is maximal}:
$$R=\frac{b T_{exp}+\sigma_e^2}{F^{SN} T_{exp}/2\pi\left[\frac{\sigma}{p}\right]^2}
=\frac{total\ noise/pixel}{peak\ flux\ of\ the\ SN/pixel}.$$
For long exposures, the electronic readout noise will always be
negligible compared to the fluctuations of the background brightness.
Then R can be written:
$$R=1.13\left[\frac{seeing}{1\ arcsec}\right]^2 10^{\frac{M-\mu}{2.5}},$$
where M is the magnitude of the SN and $\mu$ is the total background
magnitude per $arcsec^2$.
The D factor variation with R is given in Fig. \ref{fig1}~;
we easily check that for R=0 (no background), the
resolution is simply given by the gaussian fluctuation of the number
of signal photoelectrons, as expected.
The following formula is an excellent approximation (within 2\%) of D, at
least up to R=10:
$$D\simeq 0.27+1.5\sqrt{R+0.25}.$$
For a given passband $w\in \{{\bf UBVRI}\}$, $F^{SN}_w$ is related
to the $M_w$ magnitude of the SN and to the collecting
surface S of the telescope through the relation:
$$F^{SN}_w = \Phi_w(M_w=0)\times S\times 10^{-M_w/2.5},$$
where $\Phi_w(M_w=0)$ is the photoelectron flux in the w passband
per second per $m^2$ collecting surface,
for a SNIa with magnitude $M_w=0$.
Then the exposure time $T_{exp}$ needed to reach a photometric
precision $\delta$ is given by the expression:
\begin{eqnarray*}
T_{exp}&=&\frac{D^2}{F^{SN}_w\times \delta^2}\\
&=&\frac{10^{M_w/2.5}}{S\times\Phi_w(M_w=0)}\times\frac{D^2}{\delta^2}\\
&\simeq& \frac{10^{M_w/2.5}}{S\Phi_w(M_w=0)}
\frac{1}{\delta^2}
\left[0.27+1.5\sqrt{0.25+1.13\left[\frac{seeing}{1\ arcsec}\right]^2
10^{\frac{M_w-\mu_w}{2.5}}}\right]^2.
\end{eqnarray*}
This exposure time scales with the inverse of the telescope
collecting power (surface $\times$ transmission)
and with the inverse of the square of the requested resolution.
The scaling with the seeing is more complex, and depends on the
contrast between the signal (from the SN) and the background flux.

For the MARLY telescope
$S=\frac{\pi}{4}\times(d_{prim.}^2-d_{sec.}^2)=0.57\ m^2$
where $d_{prim.}=0.98~m$ is the primary mirror diameter
and $d_{sec.}=0.49~m$ is the diameter of the obstructing secondary mirror.
Then $T_{exp}$ can be expressed as follows :
\begin{equation*}
\begin{split}
T_{exp}
& \simeq \frac{(175 s)}{\left[\delta(in \%)\right]^2}
\left[\frac{\Phi_w(M_w=0)}{10^9}\right]^{-1}\times	\\
& \quad \times 10^{\frac{M_w-17.5}{2.5}}
\left[0.27+1.59\sqrt{0.22+\left[\frac{seeing}{1\ arcsec}\right]^2
10^{\frac{M_w-\mu_w}{2.5}}}\right]^2.
\end{split}
\end{equation*}
When the background flux is negligible (i.e. if $\mu_w$ is large, or if the
SN is bright), the last factor in the expression of $T_{exp}$
is 1, and the formula can be simplified.

\end{document}